# Highly sensitive piezotronic pressure sensors based on undoped GaAs nanowire ensembles

Yonatan Calahorra[1,†], Anke Husmann[1,†], Alice Bourdelain[2], Wonjong Kim[3], Jelena Vukajlovic-Plestina[3], Chess Boughey[1], Qingshen Jing[1], Anna Fontcuberta i Morral[3,4], Sohini Kar-Narayan[1]*

[1] *Department of Materials Science and Metallurgy, University of Cambridge, 27 Charles Babbage Cambridge CB3 0FS, UK*

[2] *Institut des Nanotechnologies de Lyon - Université de Lyon, UMR 5270 - CNRS, Ecole Centrale de Lyon, France*

[3] *Laboratory of Semiconductor Materials, Institute of Materials, School of Engineering, Ecole polytechnique fédérale de Lausanne, Lausanne, Switzerland*

[4] *Institute of Physics, School of Basic Sciences, Ecole polytechnique fédérale de Lausanne, Lausanne, Switzerland*

[†] *These authors contributed equally to this work*
*Email: sk568@cam.ac.uk

## Abstract

Semiconducting piezoelectric materials have attracted considerable interest due to their central role in the emerging field of piezotronics, where the development of a piezo-potential in response to stress or strain can be used to tune the band structure of the semiconductor, and hence its electronic properties. This coupling between piezoelectricity and semiconducting properties can be readily exploited for force or pressure sensing using nanowires, where the geometry and unclamped nature of nanowires render them particularly sensitive to small forces. At the same time, piezoelectricity is known to manifest more strongly in nanowires of certain semiconductors. Here, we report the design and fabrication of highly sensitive piezotronic pressure sensors based on GaAs nanowire ensemble sandwiched between two electrodes in a back-to-back diode configuration. We analyse the current-voltage characteristics of these nanowire-based devices in response to mechanical loading in light of the corresponding changes to the device band structure. We observe a high piezotronic sensitivity to pressure, of ~7800 meV/MPa. We attribute this high sensitivity to the nanowires being fully depleted due to the lack of doping, as well as due to geometrical pressure focusing and current funnelling through polar interfaces.

**Introduction**

Nanoscale piezoelectric semiconductors have attracted considerable attention in the past decade for applications in sensing, optics and energy harvesting [1-3]. This stems from the unique set of properties offered by the combination of piezoelectricity and semiconductor physics. For example, the piezoelectric effect, where mechanically induced changes to atomic structure induce electrical polarisation (and vice-versa) [4], is transient by nature. In other words, applying pressure to a piezoelectric crystal will yield a voltage (or current) peak, but not act as a DC source. In that sense, a piezoelectric pressure sensor is actually a pressure-change sensor. The underlying reason is the retention of charge neutrality. Conversely, in semiconductors, potentially there are regions of high and low conductivity in close proximity, such as junction-related depletion regions [5]. Consider a semiconductor which is also piezoelectric, *e.g.* III-V, III-N or II-VI materials, that forms a p-n or metal-semiconductor junction. If the depletion region become strained, then the corresponding change in polarisation will result in a change in surface charge and hence surface potential, effectively resulting in a change in junction barrier height. This effect is not time-dependent since the polarisation charges would not be neutralised. This effect has been coined as the piezotronic effect [3], and it bears potential in mechanical sensing [1], mechanically controlled logic [6] and optics [7]. It also affects the operation of III-N based high electron mobility transistors[8], and is closely related to ferroelectric-metal contacts [9].

The current-voltage (*I-V*) characteristics of a single Schottky junction is given by [5]

$$I \sim exp\left(\frac{q\Phi_B}{k_B T}\right) exp\left(\frac{qV}{nk_B T}\right) x \left(1 - exp\left(-\frac{qV}{k_B T}\right)\right) \quad (1)$$

where $\Phi_B$ is Schottky barrier height, $V$ is the applied voltage, $n$ is the ideality factor, $k_B$ is the Boltzmann constant, $T$ is the temperature and $q$ is the electronic charge. Following Eq. (1), a simple expression for the change in current due to change in barrier height is given by

$$ln(I_{Stress}/I_0) \sim -\Delta\Phi_B \left(\frac{q}{k_B T}\right) \quad (2)$$

The current through back-to-back Schottky diodes (as found in a semiconductor sandwiched between two metallic electrodes) is more complex, since the applied voltage is dropped across both junctions, as well as the semiconductor, taking continuity into consideration. The *I-V* characteristics in this case are governed by the semiconductor depletion, and flat-band conditions [10].

Most work on nanowire (NW) piezoelectricity and piezotronics has been devoted to III-Ns and ZnO, which are the most commonly known piezoelectric semiconductors [6, 11-16]. This is intriguing, considering that the initial observations of mechanical effects on Schottky barrier height were obtained on III-V materials [17-20]. The piezotronic effect in (non-nitride) III-V materials is important due to their applications in high-speed electronics and optoelectronics [21], and in NWs due to potential integration with silicon [22, 23]. However, there are not many reports on piezoelectricity and piezotronic effects in III-V NWs [24-28], although Ref. [29] provides a comprehensive review on the subject. In previous work from our group, we have used non-destructive piezoresponse force microscopy (PFM) [30] to examine the converse piezoelectric effect in GaAs and InP NWs [28]. Others have also focused on GaAs NWs, studying piezoelectric generation of NW ensembles [26], as well as single NW current generation and piezotronics using an atomic force microscopy (AFM) apparatus [27]. In this context, it is useful to examine the electro-mechanical effects in a practical device configuration. Here, we demonstrate and analyse the piezotronic effect in Schottky diodes based on GaAs NW ensembles. We consider the complex nature of device contacts, and qualitatively determine the stress distribution in the NWs through analysis of current response to mechanical loading in a back-to-back diode configuration.

**Materials and Methods**

Piezotronic pressure-sensitive devices were fabricated by sandwiching GaAs nanowire ensembles between two electrodes. GaAs nanowires have been grown through molecular beam epitaxy (MBE) by Ga-assisted self-catalysed method on Si substrates covered with a thin oxide layer. Scanning electron microscopy (SEM, Hitachi TM3030 was used to characterise NW morphology and monitor device processing. Bottom left SEM in Figure 1 shows the growth results, comprising a mix of zinc-blende (ZB) NWs, wurtzite (WZ) NWs, and a parasitic growth (see Supporting Information S1, for full details).

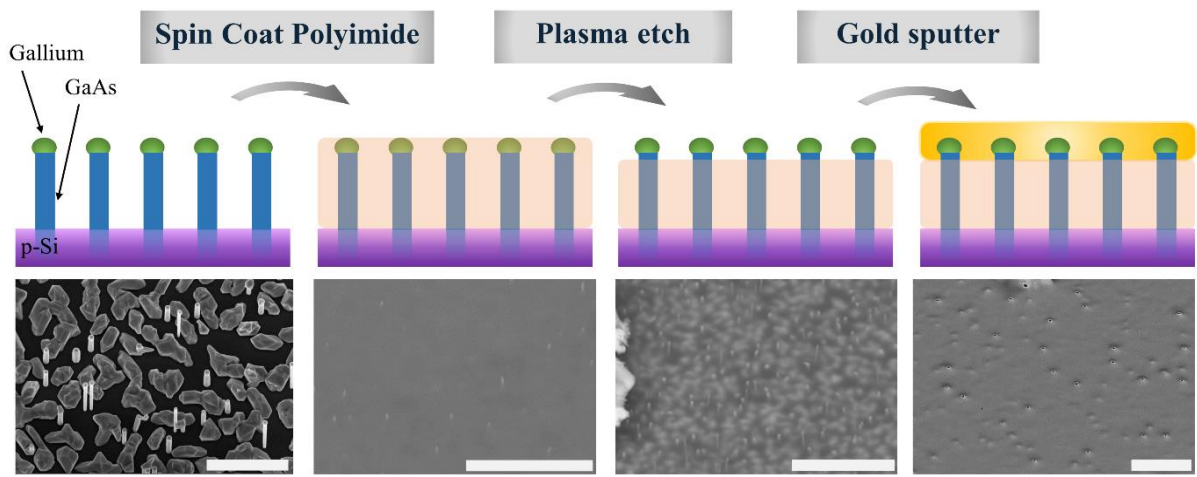

Figure 1. Schematic of device fabrication process, and corresponding SEM images: the silicon substrate bearing grown GaAs NWs is spin coated with polyimide, etched and sputtered with gold. Left scale bar is 1.5 μm, all other scale bars are 5 μm.

Piezotronic devices were fabricated using these nanowire ensembles as follows (see Figure 1): polyimide was spin coated as a spacer layer, then etched back to expose the NW tips, followed by the devices being sputtered with gold as the top electrode(see Supporting Information S1 for full details).

The p-Si substrate served as the bottom electrode and was connected to a conductive substrate via silver paint for electrical access. Figure 1 shows a schematic of the device fabrication and corresponding SEM images of the different fabrication stages. The area of a single device was roughly 0.1 cm$^2$, see also Figure 2a inset. Considering the back etching step of the spacer layer, we assume the contacts are mostly formed to the longest NWs, usually associated with ZB growth. We did not consider the mechanical or electrical effects of uncontacted material.

*I-V* Measurements were performed under different loading scenarios using LabView® to control a Keithley 2400 SourceMeter. Applied voltages were swept from 0V -> 1V -> -1V -> 0V with a sweep rate of 0.1 V/s while the current was measured in a two-point configuration. Devices were loaded with a set of calibration weights. The load was sequentially increased, and a comparative voltage sweep was done at the end of each run to ensure that the devices recover their original unloaded *I-V* characteristics.

**Results and Discussion**

Figure 2a shows the *I-V* measurements of a typical GaAs NW ensemble device, before, during and after mechanical loading. The curves changed and current was found to diminish (in absolute value) with the application of compressive stress, i.e. when a weight was placed on top of the device. Noticeably, this happened for both voltage polarities, unlike some piezotronic devices reported in the literature, where current on one side increases while it decreases on the other side [31]. There was, however, a difference in the observed current response for the two voltage polarities: the change was significantly larger when negative voltage was applied, compared to when a positive voltage was applied. The negative polarity voltage sweep also gave rise to a higher current. This trend was observed in an additional device (See SI Figure S2).

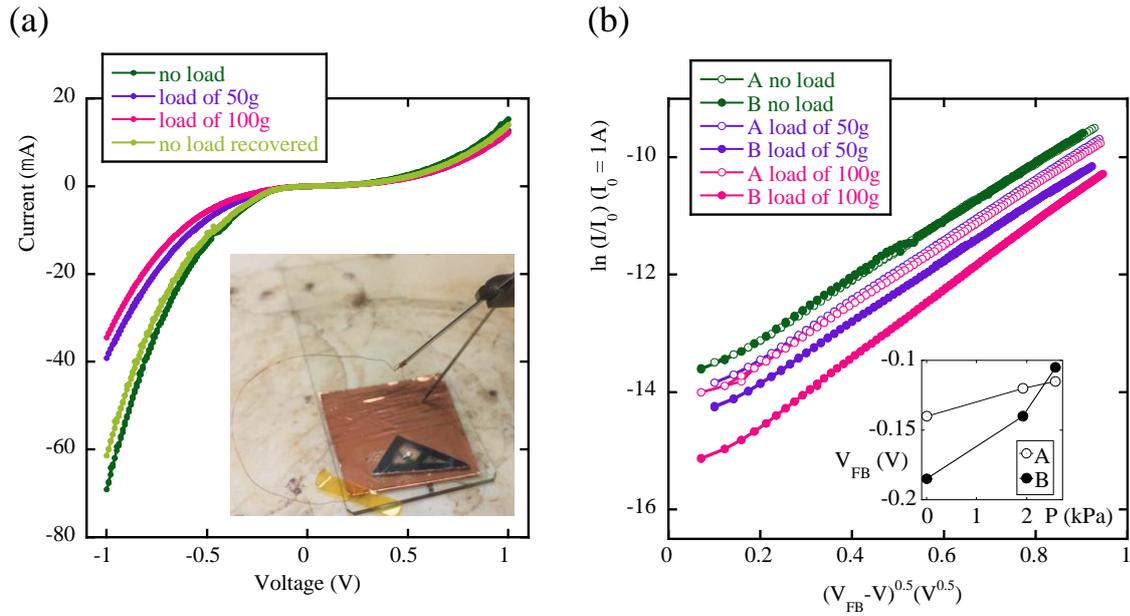

Figure 2: (a) *I-V* curve for Device B under load. Inset: device mounted for measurements (b) $ln(I) \sim \sqrt{V_{FB} - V}$ for two devices for the negative polarity of the *I-V* curves. Inset: Values for $V_{FB}$.

We start by discussing these findings qualitatively: the GaAs NWs are sandwiched between a gold electrode at the top and highly conducting p-type silicon at the bottom, and the shape of the *I-V* curve obtained agrees with back-to-back (BtB) diode characteristics. Generally, in a BtB configuration, for any given applied voltage, one diode is forward biased while the other is reverse biased, thereby the latter accounts for most of the voltage drop - and is the limiting element for the electronic current. In our device, voltage was applied to the top gold electrode while the p-Si substrate was grounded. A positive voltage corresponds to the substrate/NW junction being reversed biased, while the metal/NW junction current determined by the actual potential drop on the junction. The smaller current response to mechanical loading in this voltage polarity indicates that while the current is dictated by the substrate/NW junction, it remains unaffected by the stress, therefore indicating that this junction and the NW segments closest to it undergo little or no deformation. This result is further discussed in detail below.

In negative polarity, the substrate/NW junction is forward biased.. It is useful to examine the device in better detail. In BtB Schottky diodes, a large enough reverse bias may also correspond to injection of minority carriers [10] (electrons in our case – considering the p-Si as determining the majority carriers). Our device is composed of a semiconducting junction and a Schottky junction, and therefore, in the negative polarity, it is possible that the substrate/NW junction injects majority carriers (holes), and the metal/NW junction injects electrons. We apply the results obtained by Sze [10] to examine the operation of our device in negative polarity. In our case, due to the nominally intrinsic nature of the NWs, and their inherent size dependent properties favouring depletion[32, 33], it is reasonable to expect the NW to be depleted at all voltages, and the current to obey the corresponding relations, i.e. $ln(I) \sim (V - V_{FB})^2$ or $ln(I) \sim \sqrt{V - V_{FB}}$ [10], depending on whether the device is beyond application of the flat-band voltage, $V_{FB}$. Figure 2b shows the graphs for the measured current for $ln(I) \sim \sqrt{V - V_{FB}}$, which provided a better fit than $ln(I) \sim (V - V_{FB})^2$ (see the second fit in Supporting Information Figure S3). This suggests that for most of the applied voltages the device is beyond flat-band conditions. This fitting also allows the extraction of $V_{FB}$ as a function of applied compression (Figure 2b inset), showing that $V_{FB}$ is reduced (in absolute value) almost linearly with compression. Next we analyse the band diagram of the device to examine our understanding of its operation and result.

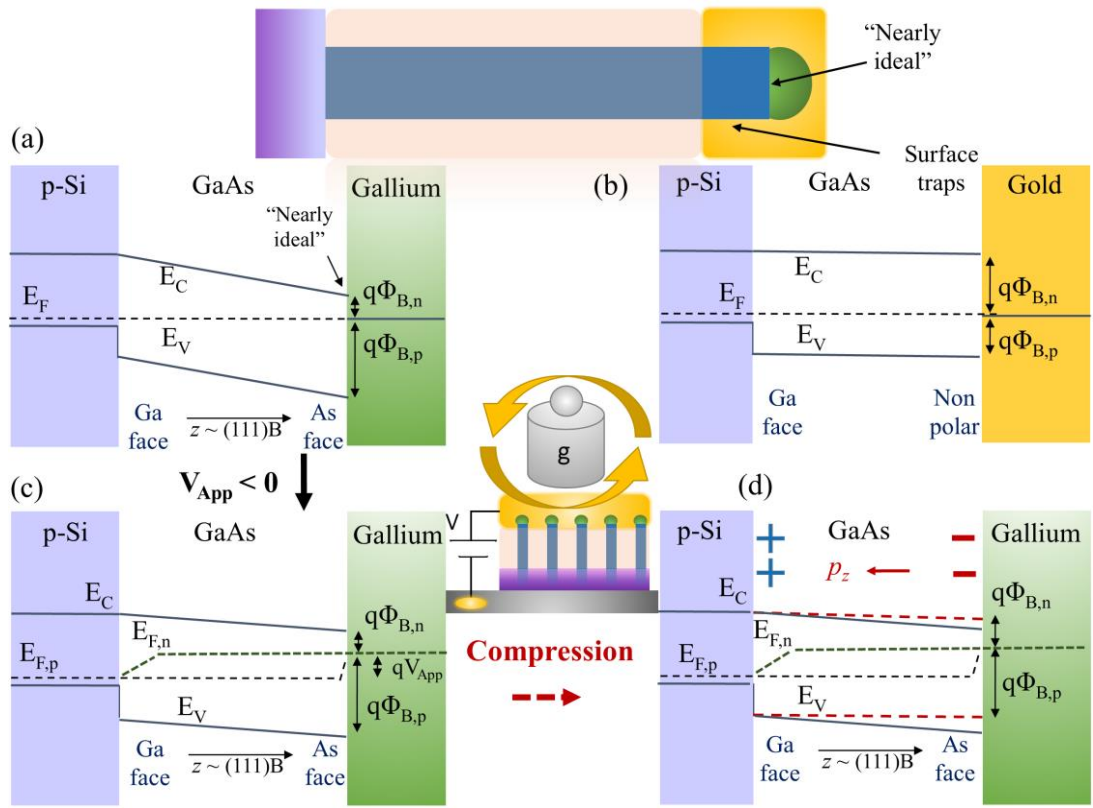

Figure 3. Schematic band diagrams of the NW device (shown in the upper inset. a) Unbiased, uncompressed axial band diagram - gallium contact; b) Unbiased, uncompressed band diagram corresponding to the sidewall contact; c) axial band diagram under negative bias; d) axial band diagram under negative bias and compression (shown schematically in the middle inset).

Figures 3a-d show the band diagrams corresponding to equilibrium, negative polarity, and the addition of compressive stress. This is a schematic analysis based on the conditions in the GaAs boundaries. Specifically, considering the depleted nature of the device as mentioned above, we examine two possible contact configurations for the metal/NW junction, due to contact potentially forming

between the NW growth front and the gallium catalyst, or the NW sidewalls and the deposited gold electrode. Figures 3a and b are plotted considering these two effects: for the growth interface, the Si/GaAs pn junction, we consider ideal band continuity, thus determined by the similar electron affinities of silicon and GaAs (~ 4.05 eV) [5]; next, the conduction and valence bands of the GaAs NW are mostly linear due to depletion (Poisson equation becomes Laplace equation following the lack of ionised dopants); finally we consider a "nearly ideal" interface between the gallium droplet (work function of ~ 4.2 eV) and the (111)B facet of the NW, formed in-situ (Figure 3a), and a post-processing interface between the {1-10} or {11-2} facets of the NW sidewalls and the gold, dominated by charge traps, with Fermi level pinning to ⅓ of the band gap [10].

Let us now consider the two voltage polarities and the two contact configurations. Clearly, the gold contact favours hole injection (in positive applied voltage polarity), while the gallium contact favours electron injection (in negative applied bias polarity). It is therefore possible that during polarity change, the active contact also changes. Interestingly, the gold contact is formed on non-polar GaAs surfaces, which agrees with the lower electromechanical response corresponding to the positive polarity. Another factor which probably contributes to this finding is the mechanical encapsulation of the NWs, effectively protecting the substrate/NW junction, thereby further reducing the electromechanical response. Notably, plotting $ln(I) \sim \sqrt{V - V_{FB}}$ for the forward bias yields a very small value (~ $k_B T$, Supporting Information Figure S4) for $V_{FB}$. This aligns very well with the gold-contact band diagram (Figure 3b) and further validates our understanding of the device operation. We continue analysing the device assuming that the gallium contact, formed on the polar (111)B face of the NW, is active in the negative polarity.

Figure 3c depicts the gallium contact under negative voltage, injecting minority carrier electrons, while the substrate/NW Pn junction is forward biased (injecting holes), resulting in quasi Fermi levels for holes and electrons, that are plotted schematically. Figure 3d depicts the device upon application of compression. The axial piezoelectric coefficient ($d_{33}$ or $e_{33}$) of (111) oriented GaAs NWs is negative and therefore upon application of compressive stress, a negative polarisation charge develops at the arsenic face, and positive charge at the gallium face [28, 34]. Overall, the polarisation is oriented along (111)A, resulting in upward bending of the bands at the arsenic face, thus effectively increasing the barrier for electrons, and reducing the current. Another outcome of the mechanically induced band shift is that the bands are pushed towards flat-band conditions, meaning that a smaller bias is required in order to reach flat-band conditions. This is in perfect agreement with the fitting results of the devices shown in the inset of Figure 2b, where $V_{FB}$ decreases (in absolute value) with compression.

We now discuss the results in better detail in light of our understanding of the device structure and corresponding band diagrams. Figure 4 shows the calculated gauge factor, GF, defined as the relative change in resistivity normalised by strain. This calculation is an estimate, considering we did not measure the actual compression in the NWs, rather we only have values of the applied pressure. In order to assess strain, we consider the interaction to be elastic, and neglect any plastic deformation of the contacts. This is justified based on the successful retention of the unloaded *I-V* characteristics. We use the following values for Young's modulus: $Y_{GaAs,111}$ = 130 GPa [35], and $Y_{polyimide}$ = 2.5 GPa (from Dupont technical sheets), and we consider a simplified polyimide-GaAs NW composite, comprising ~1% GaAs, giving an overall effective elastic modulus of 3.75 GPa. Using this value to estimate strain for a given load, our GaAs NW piezotronic devices were found to be highly sensitive to pressure with a peak GF of ~$10^5$. This is considerably higher than piezoresistive devices with GF of ~ 200, and also compared to piezotronic devices based on ZnO [11, 31] or InAs NWs (~1000)[36].

For a quantitative comparison, we examine the known quantities used, i.e. weight and associated pressure, and examine the piezotronic pressure sensitivity, $\Delta\Phi_B/\Delta P$, where *P* is the applied pressure. We compare our results to Ref. [11],[15],[20] and [37]. For our device, we plot the barrier height change for the weights examined (Supporting Information Table S5) and find a slope of about 7790 meV/MPa for our higher performing device (B). This suggests that our devices exhibit excellent sensitivity for pressure, the highest reported to date to the best of our knowledge. In Ref. [11] a single ZnO microwire was strained (and compressed) in ratios up to ±1%, with a corresponding change in barrier height of ±5 meV. To convert this strain to stress, we take $Y_{ZnO,wire}$ = 80 GPa, which is a value somewhat smaller than accepted for bulk [38]. Therefore, Zhou *et al.* observed a pressure sensitivity of 5/800 = 0.0063 meV/MPa in their study. Kiel *et al.* [15] measured 9/65 = 0.14 meV/MPa sensitivity for a single crystal ZnO Schottky barrier. More recently, Wang *et al.* [37] reported a record high piezotronic pressure sensitivity of 1400 meV/MPa, by combining two mirrored ZnO platelets, thus taking advantage of the surface polarisation effect on both device contacts, the sensitivity of this configuration was further studied by Kiel *et al.* [39]. If so, our result of 7790 meV/GPa is remarkable considering that ZnO is about an order of magnitude more piezoelectric than GaAs [28]. It is also striking when compared to the only study we know of piezotronic sensitivity in bulk GaAs, where Chung *et al.* measured 0.0067 meV/MPa for the (111) direction [20], indicating a 6 orders of magnitude increase in sensitivity of the NW device compared to bulk diodes.

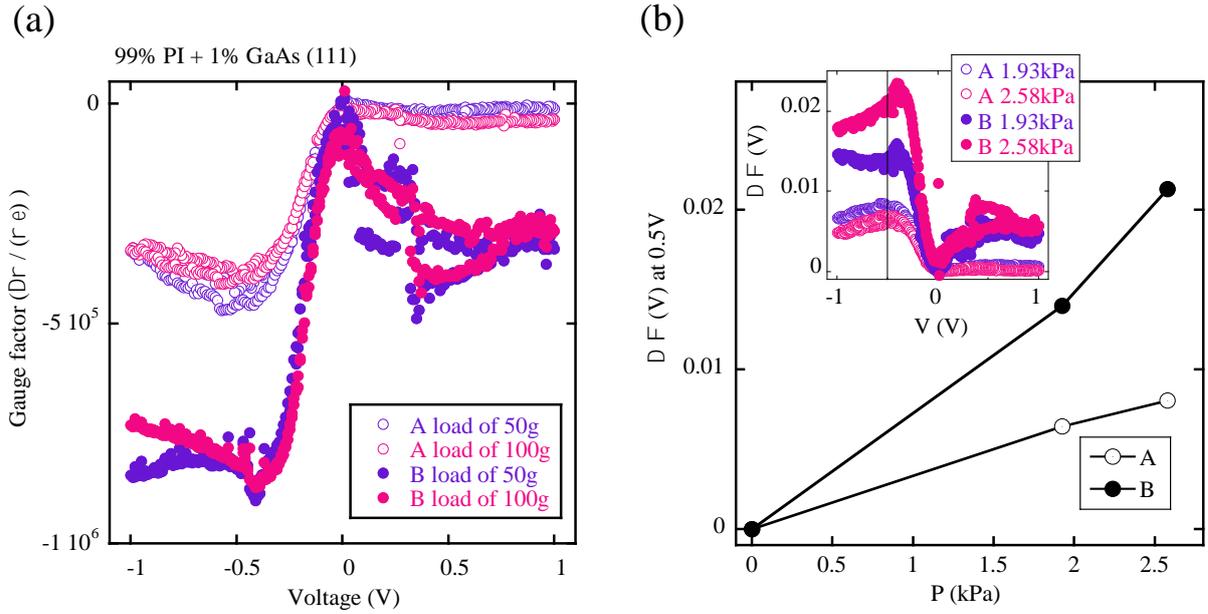

Figure 4: (a) Gauge factors for two devices under two different loading conditions. (b) extracted $\Delta\Phi_B$ as a function of pressure at -0.5 V, with and in inset showing the variation of $\Delta\Phi_B$ as a function of voltage.

There are three possible reasons for the observed high sensitivity, the first being mechanical stress focusing, as reported for GaAs NW piezoelectric sensors [26]. In Ref. [26], the authors have shown that electrically contacting NW ensembles may result in a significant increase in effective strain in the NWs, leading to an enhancement of the apparent piezoelectric coefficients by up to 2 orders of magnitude. It is very likely that this effect significantly affects the response of our pressure-sensitive piezotronic device. Secondly, the fact that the device is completely depleted contributes to the reduction of any screening effect, thus maximising the piezotronic response. As mentioned above in the initial discussion of the piezotronic effect, the existence of the depletion region allows the existence of the non-decaying piezotronic effect to begin with. In the case where the entire length of the NW is depleted, the effect is obviously expected to be enhanced. This explanation is supported by our finding that devices with higher initial currents exhibited lower sensitivities (see the comparison between devices A and B in Figure 4). Interestingly, in a recent theoretical work, a piezotronic bipolar transistor was examined [40], where it was shown that as the transistor was strained, the dominant piezo-potential developed at the base of the transistor, i.e. the area with lowest doping that was essentially depleted. The theoretically derived GF was reported to be $\sim 10^4$, which is an order of magnitude higher than obtained by a single ZnO NW contact studied in [11]. If so, there is an established connection between depletion and enhancement of the piezotronic effect. This is of course not surprising since the piezotronic effect is defined in depletion regions. Therefore, the question of significance for a device is how much strain is transferred to areas which are depleted. Finally, another factor that might contribute to the sensitivity of our devices is the contact configuration. In our discussion we concluded that the dominant contact under negative bias is the gallium droplet, which interfaces the polar (111)B facet. Therefore, arguably, all the electrons injected into the device in negative bias "feel" the pressure. In many other studies involving single ZnO NWs, contacts are made in a wrapping configuration, where significant contact is made to non-polar facets, effectively lowering the responsivity. This aligns well with our observation of the lower response in the positive bias where contact is made through the wrapping gold. In fact if we examine $V_{FB}$ for the positive bias, we find that it does not change with pressure (Supporting Information Figure S4), indicating the lowered piezotronic response.

**Conclusions**

To summarise, we have fabricated highly sensitive piezotronic sensors using MBE-grown GaAs nanowire ensembles in a back-to-back diode geometry. We studied the current-voltage characteristics of these devices with and without mechanical loading, and found that the current response was dependent on the polarity of the applied voltage. This asymmetry in the response arose due to the inherently different nature of the Schottky diodes formed at the respective electrode/nanowire interfaces. The doped p-Si substrate on which the nanowire ensemble was grown, served as the bottom electrode was, while the top electrode comprised of sputtered gold / gallium droplet. The effect of loading on the current-voltage characteristics could be explained by considering the band diagram of the device, whereby the applied mechanical pressure resulted in polarisation changes at the surface, thus leading to a change in the band conditions and barrier height. Our best-performing device exhibited a piezotronic pressure sensitivity of 7790 meV/MPa, which is higher than that reported in the literature for other nanowire-ensemble, single nanowire and single/bicrystal configurations piezotronic pressure sensors. The high piezotronic sensitivity of the device could be attributed to increased device depletion, funnelling of the current through polar interface, as well as mechanical stress focusing as a result of our nanowire-based device geometry. We expect that these finding will serve to guide further work into the design of highly sensitive pressure sensors.

**Acknowledgements**

This work was financially supported by a grant from the European Research Council through an ERC Starting Grant (Grant no. ERC–2014–STG–639526, NANOGEN). C.B. acknowledges funding from the EPSRC Cambridge NanoDTC, EP/G037221/1, Cambridge Philosophical Society bursary, Sir Colin Corness Bursary from Magdalene College Cambridge, and EPSRC post-Ph.D. prize. A.H acknowledges funding from the Isaac Newton Trust. WK, JVP and AFiM thank funding from H2020 through the projects INDEED and Nanoembrace.